\documentclass{iopart}
\usepackage{iopams}

\usepackage{graphicx}

\def\ket#1{|#1\rangle }
\def\bra#1{\langle#1 | }

\def\punkt{\;\; .}
\def\komma{\;\; ,}
\def\expect#1{\langle#1 \rangle}

\def\w{\omega}
\def\H{{\cal H}}
\def\e{\epsilon}

\begin{document}

\title{Dynamics of large anisotropic spin in a sub-ohmic  dissipative
  environment close to a quantum-phase transition}

\author{Frithjof B. Anders}
\address{Institut f\"ur Theoretische Physik, Universit\"at Bremen,
  P.O. Box 330 440, D-28334 Bremen, Germany}

\ead{anders@itp.uni-bremen.de}

\date{\today}

\begin{abstract}
We investigate the dynamics of a large anisotropic spin 
whose easy-axis component is coupled to a bosonic bath with a spectral
function $J(\w)\propto \omega^s$. Such a spin complex might be
realized in a single-molecular magnet. Using the
non-perturbative renormalization group, we calculate the line of quantum-phase
transitions in the sub-ohmic regime ($s<1$). These quantum-phase
transitions only occur for integer spin $J$. For half-integer $J$,
the low temperature fixed-point is identical to the fixed-point of the
spin-boson model without quantum-tunneling between the two
levels. Short-time coherent oscillations in the spin
decay prevail even into the localized phase in  the sub-ohmic
regime. The influence of the reorganization energy and the recurrence
time on the decoherence  in the absence of quantum-tunneling is discussed.
\end{abstract}

\maketitle

\tableofcontents

\section{Introduction}

Understanding the influence of the environment on the
non-equilibrium dynamics of quantum  systems remains
one of the challenging questions of theoretical physics. 
A finite number of quantum mechanical degrees of freedom,  a large
spin or a qubit interacting with a infinitely large bath of
non-interacting bosons  with a continuous energy spectrum 
represents a typical class of model examples for  such systems. Its
simplest version, the spin-boson model \cite{Leggett1987}, has contributed
tremendously to our understanding of dissipation in quantum
systems \cite{Weiss1999}.

Single-molecular magnets (SMM) based on Mn$_{12}$-actetate  or
Fe$_8$-complexes behave as large single spins at low temperatures. The
magnetic anisotropy of the molecules prevent a simple relaxation of the
spin. The resulting hysteretic behavior in the magnetization has 
been subject to intensive experimental \cite{Barbara1999,
BarbaraSmm2000,Gatteschi2003,kerenSMM2007} and 
theoretical
investigations~\cite{Stamp1998,LeuenbergerLoss1999,Fernandez2003,VorrathBrandes2005}
(see Ref.~\cite{Gatteschi2003} for an overview). Using Wilson's
numerical renormalization group (NRG)
method \cite{Wilson75,BullaCostiPruschke2007},  it was
shown \cite{RomeikeSchoeller2006} that  the combination of quantum
tunneling and antiferromagnetic coupling to a metallic substrate
induces a Kondo effect in such SMM. The recently developed
time-dependent numerical renormalization group method
(TD-NRG)\cite{AndersSchiller2005,AndersSchiller2006} has been employed
to analyze the real-time dynamics  of such large spin
\cite{RoosenTdNRG2008}.

In this paper we will investigate the equilibrium and non-equilibrium spin
dynamics of a anisotropic spin coupled to a  dissipative
bosonic bath. Such a spin complex might be realized in a
single-molecular magnet in which the 
physical properties depend on the type of coupling to the
environment. Previously, detailed expansion of the spin-lattice
relaxation \cite{PolitiMn12Relax1995}  in powers of the local spin
operator have been considered to estimate  the
spin-relaxation rates \cite{PolitiMn12Relax1995,LeuenbergerLoss1999} by Fermi's
golden-rule  arguments. Vorrath and Brandes have reduced the spin-bath
coupling to a linear term proportional to the $S_z$
operator \cite{VorrathBrandes2005}.  A Weisskopf-Wigner coupling to the
environment would describe the interaction with a fluctuating
quantized magnetic field. This might be relevant for recent dephasing
experiments \cite{kerenSMM2007} in SMM. Here, we  will restrict
ourselves to diagonally coupled spin-boson model
interaction \cite{Leggett1987,VorrathBrandes2005}.

The real-time dynamics in the spin-boson model is usually investigated
using the noninteracting-bilp
approximation \cite{Leggett1987,NesiSBMGrifoni2007}  or by Bloch-Redfield
type of approaches \cite{BlochRedfield1964}. Recently, it has been
demonstrated that the non-perturbative numerical renormalization 
group \cite{Wilson75,BullaBoson2003,BullaVoita2005,BullaCostiPruschke2007}
and its extension to non-equilibrium dynamics, the
TD-NRG \cite{AndersSchiller2005,AndersSchiller2006}, is particularly
suitable to access the  quantum-critical region of the spin-boson model
in a sub-ohmic environment \cite{BullaBoson2003,BullaVoita2005} and is able
to describe the real-time dynamics on short-time scales as well as
exponentially long-time scales \cite{AndersBullaVojta2007}.

In
order to gain a better understanding of the  non-equilibrium dynamics,
we also present the  equilibrium properties  of the model. We will map
out the quantum-critical line separating a delocalized phase in weak
coupling and a local phase at strong coupling to the dissipative
sub-ohmic bath. The continuous quantum-phase transitions found for $0<s<1$
and integer spin $J$ resemble the phase diagram previously reported
for the spin-boson model by Bulla et al.~\cite{BullaBoson2003}. No phase transition  was found for half
integer spin $J$. Information on the equilibrium dynamics is provided
by the equilibrium spin correlation function $C(\omega)$. 
The non-equilibrium spin dynamics of a  anisotropic spin complex coupled to a
dissipative bosonic bath is investigated in response to a 
sudden change of magnetic field close to quantum phase transitions.

\section{Modelling of  a large spin in a dissipative environment}

\begin{figure}
  \centering
  \begin{tabular}{cc}
  \includegraphics[width=5cm]{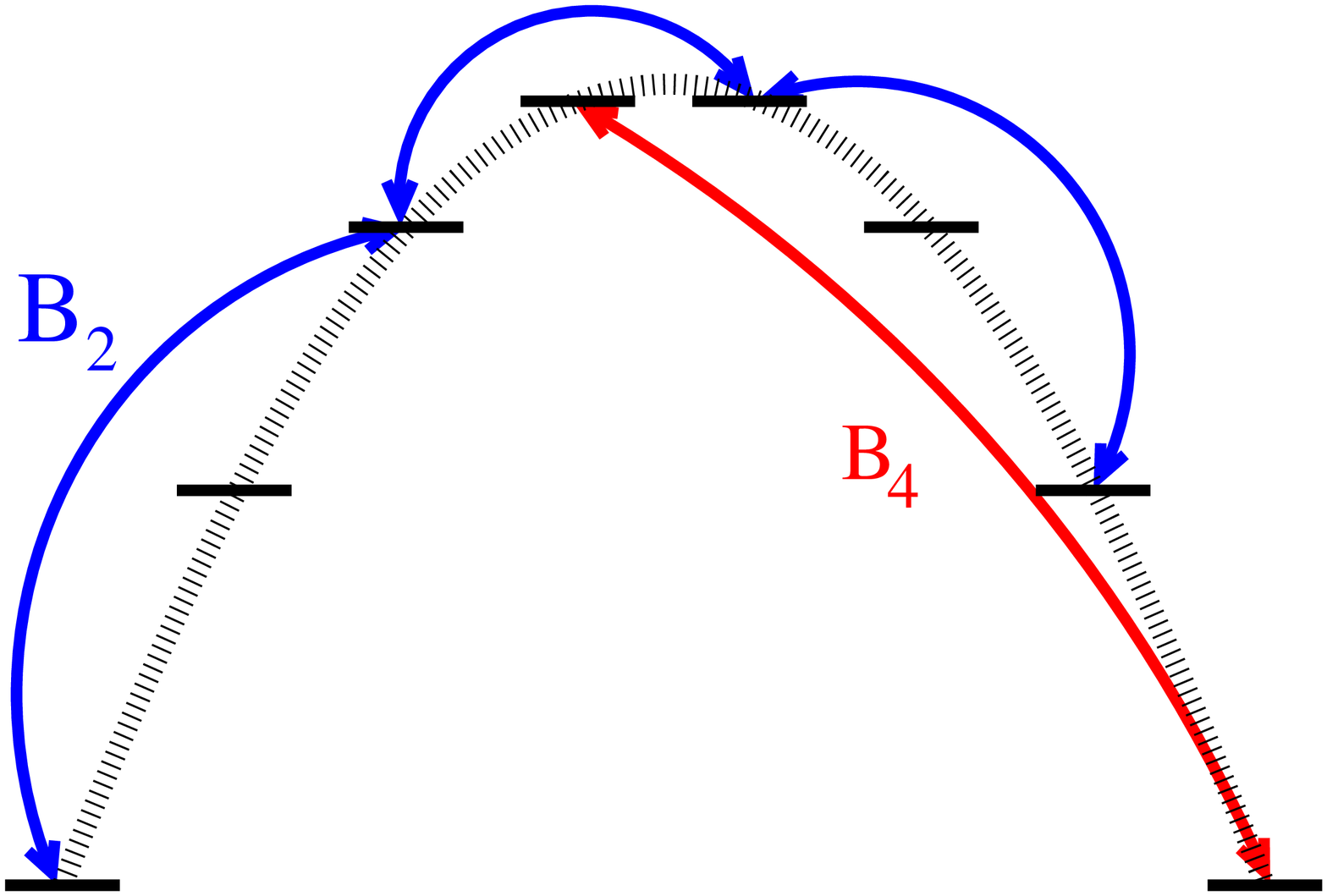} &
  \includegraphics[width=5cm]{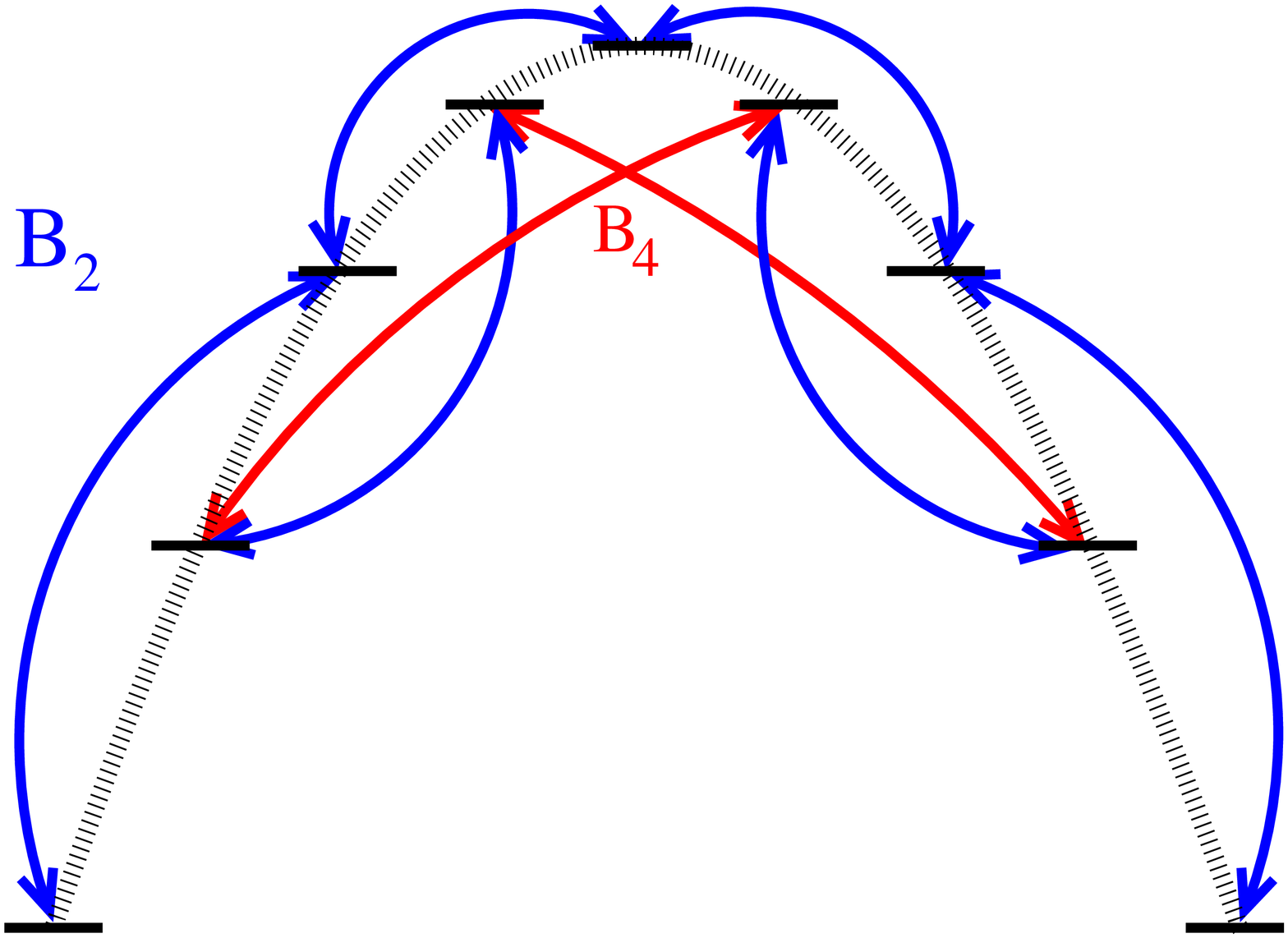}     \\
  (a) & (b)
  \end{tabular}

  \caption{Level scheme for (a) spin $J=7/2$ and (b) $J=4$.
    The two states $\ket{\pm J}$ are only connected by the quantum
    tunneling terms $B_2$ and $B_4$ for integer spin $J$.
    }
\label{fig:energy-parabel}
\end{figure}

The local Hamiltonian modeling
single-molecular magnets \cite{Gatteschi2003}
\begin{eqnarray}
\label{eqn:H}
  \H &=& \H_{loc} + \H_I + H_{bath}
\end{eqnarray}
consists of a single large spin of size  $J$ subject to an easy-axis
anisotropy energy $A$ and quantum-tunneling
terms \cite{Gatteschi2003,RomeikeSchoeller2006}  $B_{2n}$,
which induce transitions between the eigenstates of $S_z$:
\begin{eqnarray}
  \H_{loc} &=& -AS_z^2 +\sum_{n=1,2} B_{2n}\left(S_-^{2n}+S_+^{2n}\right)
\punkt
\end{eqnarray}
These  quantum-tunneling terms stem from a  reduction of the $U(1)$
easy-axis spin symmetry  to the discrete symmetry group of the SMM
molecule. Assuming $A\gg |B_{2n}|$ and a positive $A$ , the energies
of states $\ket{J_z}$ are located on an inverted energy parabola. As
depicted in Fig.~\ref{fig:energy-parabel}, there is a fundamental difference
between integer and half-integer values of the
spin \cite{Gatteschi2003,RomeikeSchoeller2006}: the two ground states
$\ket{\pm J}$ are connected via the quantum-tunneling terms $B_{2n}$
only for integer values of $J$, while for half-integer values the
ground states remain disconnected. Without further transition
mechanism, a half-integer  anisotropic spin cannot relax into 
thermodynamic ground state after switching off the spin-polarizing
external magnetic field. 

We use a generalization \cite{VorrathBrandes2005} of the  spin-boson
model (SBM) 
\begin{eqnarray}
  \H_I&=& S_z \sum_{q}\lambda_q\left(b_q^\dagger + b_{-q}\right) \\
H_{bath} &=& \sum_{q} \w_q b_q^\dagger b_q
\label{eqn:H-I}
\end{eqnarray}
to describe the coupling to a dissipative bath. Here, $ b_q^\dagger$
creates a boson in the mode $q$ with energy $\w_q$. 
The coupling between spin and bosonic bath is completely specified
by the bath spectral function\cite{Leggett1987}
\begin{eqnarray}
J(\omega) &=& \pi \sum_{q} \lambda_{q}^{2} \delta(\omega -\omega_{q})  
\punkt
\end{eqnarray}
The asymptotic low-temperature behavior is determined by the low-energy
part of the spectrum.
Discarding high-energy details, the standard parametrization is
\begin{equation}
  J(\omega) = 2\pi\, \alpha\, \omega_c^{1-s} \, \omega^s\,,~ 0<\omega<\omega_c\,,\ \ \ s>-1
\label{power}
\end{equation}
where the dimensionless parameter $\alpha$ characterizes the
dissipation strength. $\omega_c$ is a cutoff energy and is used as the
energy scale throughout the paper. The value $s=1$ corresponds to the case of ohmic dissipation. In the usual
spin-boson model (SBM), the local spin Hamiltonian
\begin{eqnarray}
  H_{loc}^{SBM} &=& -\frac{\e}{2}\sigma_z +\frac{\Delta}{2}\sigma_x
\end{eqnarray}
is parametrized by an energy splitting $\e$ and a quantum-tunneling
term $\Delta$ where $\sigma_i$ are Pauli matrices\cite{Leggett1987}.

\section{Numerical renormalization group }

\subsection{Equilibrium NRG for a bosonic environment}

In all our calculations we have used Wilson's  numerical
renormalization group (NRG) method. Wilson's NRG method is a very
powerful tool for accurately calculating  equilibrium properties of
quantum impurity models \cite{BullaCostiPruschke2007}. Originally the
NRG was invented by Wilson for a fermionic bath to solve the Kondo
problem \cite{Wilson75,KrishWilWilson80a}.  The method was recently
extended to  treat quantum impurities  coupled to a bosonic bath
\cite{BullaBoson2003,BullaVoita2005}, combination of fermionic and
bosonic baths \cite{GlossopIngersent2007} and real-time dynamics out
of equilibrium
\cite{Costi97,AndersSchiller2005,AndersSchiller2006}. The
non-perturbative NRG approach has been successfully applied to
arbitrary electron-bath couplings
strength
\cite{BullaBoson2003,BullaVoita2005,GlossopIngersent2007,AndersBullaVojta2007}.

At the heart of this approach is a logarithmic discretization of the
continuous bath, controlled by the discretization parameter $\Lambda >
1$; the continuum limit is recovered for $\Lambda \to 1$. Using an
appropriate unitary transformation \cite{Wilson75}, the Hamiltonian is
mapped onto a semi-infinite tight-binding chain, defined by a sequence
of finite-size Hamiltonians $\H_m$  with the impurity coupled to the
open end. The tight-binding parameters $t_m$  linking consecutive
sites of the chain $m$ and $m+1$ falls off  exponentially as $t_m \sim
\Lambda^{-m}$.  Each bosonic chain link  is  viewed as representative
of an energy shell since its energy $w_m$ also decreases as $w_m \sim
\Lambda^{-m}$ establishing an energy hierarchy.  Both ensure that
mode coupling can only occur between neighboring energy shells, which
is essential for the application  of the renormalization group
procedure. To this end, the renormalization  group transformation $R[H]$ reads
\begin{eqnarray}
\H_{m+1} &=& R [\H_m]
\nonumber \\
&=&\Lambda \H_m+t_m \left( a_m^{\dag}
a_{m+1}+a_{m+1}^{\dag} a_{m} \right)
 + w_m a^\dagger_{m+1} a_{m+1} \ ,\label{eqn:nrg-trafo}
\end{eqnarray}
where $\H_m$ is the Hamiltonian of a finite chain up to the site
$m$. The  annihilation (creation) 
operators of site $m$ are denoted by $a_m$ ($a_m^{\dag}$) and $w_m$
represents the energy of the bosonic mode of site $m$. Note that the
rescaling of the Hamiltonian $\H_m$ by $\Lambda$ ensures the
invariance of the energy spectrum of fixed point Hamiltonians under
the RG transformation $R [\H_m]$. For a detailed review on
this method see Ref.~\cite{BullaCostiPruschke2007}.

The RG transformation (\ref{eqn:nrg-trafo}) is used to set up and
diagonalize iteratively the sequence of Hamiltonians $H_m$. In the
first step, only the large spin coupling to
the single bosonic site $m=0$ is considered. It turns out to be
sufficient \cite{BullaBoson2003,BullaVoita2005,BullaCostiPruschke2007} to include only the $N_b$ lowest lying
bosonic states, where $N_b$ takes typical values of $8-12$. The
reason for that is a quite subtle one: the coupling between different
chain links decays exponentially and is restricted to
nearest-neighbor coupling by construction, both essential for the RG
procedure. In each successive step (i) a finite number of  $N_b$
bosonic states of the next  site $m+1$ are added, (ii) the
Hamiltonian matrices are diagonalized and (iii) only the lowest
$N_s$ states are retained in each iteration.  The discarding of high
energy states is justified by the Boltzmannian form of the
equilibrium density operator when  the temperature is
lowered simultaneously in each iteration step  to the order $T_m\propto
\Lambda^{-m}w_c$.

Denoting the set of low-lying eigenstates by $|r\rangle_N$ and the
corresponding eigenvalues $E_r(N) \propto O(1)$ at iteration $N$,
the equilibrium  density matrix $\rho_0$ is given \cite{BullaCostiPruschke2007} by
\begin{eqnarray}
  \hat \rho_0 &=& \frac{1}{Z_N} \sum_{r} e^{-\bar \beta E^N_r} |r\rangle_N
  {}_N\langle r| \ ,
\label{eqn:rho0}
\end{eqnarray}
where $Z_N = \sum_r e^{-\bar \beta E^N_r}$ and $\bar\beta$ are of
the order $O(1)$, such that $T_N = w_c \Lambda^{-N}/\bar \beta$. The
thermodynamic expectation value of each local observable $\hat O$ is
accessible at each temperature $T_N$ by the trace
\begin{eqnarray}
  \langle \hat O \rangle_{\rm eq} &=& \mbox{Tr}\left[\hat \rho_0 \hat
    O\right]
= \frac{1}{Z_N} \sum_{r} e^{-\bar \beta E^N_r} {}_N\langle r| \hat O
|r\rangle_N
 \label{ddT}
\, .
\end{eqnarray}
The procedure described above turns out to be very accurate because
the coupling $t_m$ between the bosonic sites along the chain are
falling off exponentially so that  the rest of the semi-infinite
chain contributes only perturbatively
\cite{Wilson75,BullaCostiPruschke2007} at each iteration $m$, while
contributions from the discarded high-energy states are exponentially
suppressed by the Boltzmann factor.

\subsection{Time-dependent NRG}

While the equilibrium properties are fully determined by the energy
spectrum of the Hamiltonian, the non-equilibrium dynamics requires
two conditions: the initial condition encoded in the many-body
density operator $\hat \rho_0$ and the Hamiltonian $\H^{f}$ which
governs its time-evolution. For a time-independent Hamiltonian, the
density operator evolves according to $\hat{\rho} (t>0)=e^{-i\H^{\rm
f} t} \hat \rho_0 e^{i\H^{\rm f} t}$. All time-dependent expectation
values $ \langle \hat O \rangle(t)$ are given by
\begin{eqnarray}
  \langle \hat O \rangle(t) &=& \mbox{Tr}\left[\hat \rho(t) \hat
    O\right]
= \mbox{Tr}\left[e^{-i\H^{\rm f} t} \hat \rho_0 e^{i\H^{\rm f} t} \hat
O\right] \;  \label{eqn:o-real-time}
\end{eqnarray}
where set $\hbar=1$.

We obtain the density operator $ \hat \rho_0$ from an independent NRG run
using a suitable initial Hamiltonian $\H^{\rm i}$. For instance,
by choosing a large local magnetic field in $\H^{\rm i}$, we can
prepare the system such that $S_z$ is fully polarized. To investigate
decoherence we choose $\H_{loc}^i$ such that 
\begin{equation}
  \ket{s} = \frac{1}{2J+1}\sum_m \ket{m}
\end{equation}
is an eigenstate of an appropriate $\H_{loc}^i$ with the lowest eigenenergy.

In general, the initial density operator $\hat \rho_0$ contains states
which are most likely superpositions of excited states of $\H^{\rm
f}$. For the calculation of the real-time dynamics of the large spin
it is therefore not sufficient to take 
into account  only the retained states of the Hamiltonian $\H^{\rm
f}$ obtained from an NRG procedure. The recently developed
time-dependent NRG (TD-NRG) \cite{AndersSchiller2005,AndersSchiller2006} circumvents this problem
by including contributions from all states. It turns out that the
set of all discarded states eliminated during the  NRG procedure
form a complete basis set\cite{AndersSchiller2005,AndersSchiller2006} of the Wilson chain which is
also an approximate eigenbasis of the Hamiltonian. Using this
complete basis, it was shown\cite{AndersSchiller2005,AndersSchiller2006} that
Eq.~(\ref{eqn:o-real-time}) transforms into the central equation of
the TD-NRG for the temperature $T_N$
\begin{eqnarray}
\langle \hat{O} \rangle (t) &=&
        \sum_{m = 0}^{N}\sum_{r,s}^{\rm trun} \;
        e^{i(E_{r}^m -E_{s}^m)t}
        O_{r,s}^m \rho^{\rm red}_{s,r}(m) \; ,
\label{eqn:time-evolution}
\end{eqnarray}
where $O_{r,s}^m = \langle r;m| \hat{O}| s;m \rangle$ are the matrix
elements of any operator $ \hat{O}$ of the electronic subsystem at
iteration $m$, and $E_{r}^m,E_{s}^m$ are the eigenenergies of the
eigenstates $|r;  m\rangle$  and $|s;  m\rangle$ of $\H^{\rm f}_m$.
At each iteration $m$, the chain is formally partitioned into a
``system'' part on which the Hamiltonian $\H_m$ acts exclusively and
an environment part formed by the bosonic sites $m+1$ to $N$.
Tracing out these environmental degrees of freedom $e$ yields  the
reduced density matrix\cite{AndersSchiller2005,AndersSchiller2006}
\begin{equation}
\rho^{\rm red}_{s,r}(m) = \sum_{e}
          \langle s,e;m|\hat \rho_{0} |r,e;m \rangle
\label{eqn:reduced-dm-def}
\end{equation}
at iteration $m$, where $ \hat \rho_0$ is given by (\ref{eqn:rho0}) using
$\H^{\rm i}$. The restricted sum $\sum^{\rm trun}_{r,s}$ in
Eq.~(\ref{eqn:time-evolution}) implies that at least one of the
states $r$ and $s$ is discarded at iteration $m$. Excitations
involving only kept states contribute at a later iteration and must
be excluded from the sum.

As a consequence,  {\em all} energy shells $m$ contribute to the
time evolution: the short time dynamics is governed by the high
energy states while the long time behavior is determined by the low
lying excitations. Dephasing and dissipation is encoded in the phase
factors $ e^{i(E_{r}^m -E_{s}^m)t}$ as well as the reduced density
matrix $\rho^{\rm red}_{s,r}(m)$.

Discretization of the bath continuum will lead to finite-size
oscillations of the real-time dynamics around the continuum solution
and deviations of expectation values from the true equilibrium at
long time scales. In order to separate the unphysical finite-size
oscillations from the true continuum behavior, we average over
different bath discretization schemes using Oliveira's z-averaging
(for details see
Refs.~\cite{YoshidaWithakerOliveira1990,AndersSchiller2006}). We
average over $N_z=16$
different bath discretizations in our calculation.

Previously, the TD-NRG has been successfully applied to the simple
spin-boson model \cite{AndersSchiller2006,AndersBullaVojta2007}, the
SMM coupled to a fermionic bath \cite{RoosenTdNRG2008} and
electron-transfer in a dissipative
environment \cite{TornowBullaAndersNitzan2008}.


\section{Equilibrium and quantum-phase transitions}
\label{equi}

\subsection{Integer Spin $J$ and quantum-phase transitions}
\label{sec:qpt}

\begin{figure}
  \centering
  \includegraphics[width=12cm,clip]{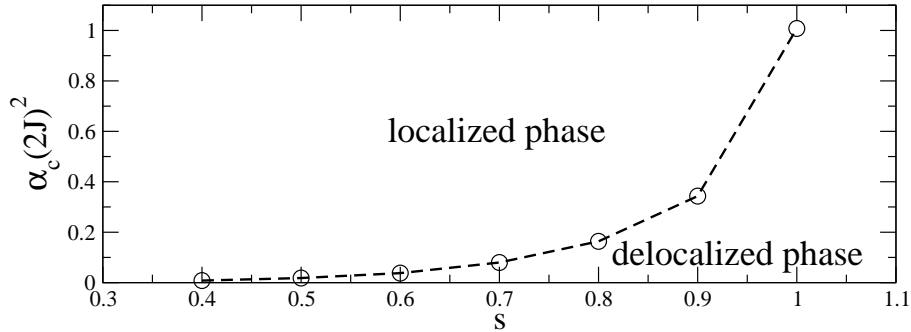}
  \caption{Zero temperature phase diagram in the sub-ohmic regime for
    J=3. The critical     coupling strength $\alpha_c(2J)^2$ is plotted vs
    the exponent s in $J(\w)$ for $A=-0.1$, $B_2=B_4=10^{-3}$,
    $H=0$. NRG Parameters: $\Lambda=2$, $N_s=100$, $N_b=8$.}
\label{fig:phase-diagram}
\end{figure}

In Fig.~\ref{fig:phase-diagram}, we plot the zero-temperature phase
diagram in the sub-ohmic regime for a moderate value of an integer spin
($J=3$.) Depicted is the rescaled critical
coupling strength $\alpha_c (2J)^2$ as a function of the bath power-law
exponent in the sub-ohmic regime. The line marks the quantum-phase
transition (QPT) between a localized and a delocalized phase. We have
obtained the critical coupling strength by investigating the NRG level
flow which shows that the QPT is associated  with a distincted fixed
point. For $s>1$, no QPT exist \cite{BullaBoson2003,BullaVoita2005}.

Analyzing the  NRG fixed-point spectra \cite{Wilson75}, 
the same qualitative picture  emerges for all integer values of
$J$. We noted that the critical 
coupling strength depends (i) on the value of the spin $J$ and (ii) the
values of the quantum tunneling rates $B_{2n}$. Only for $s\to 1$, the
critical coupling becomes independent of $B_{2n}$ and always approaches
\begin{eqnarray}
 \lim_{s\to 1} \alpha_c(s) &\approx& 1/(2J)^2 \komma
\end{eqnarray}
which can be understood from the analytical form of $\H_I$.

The fixed-point spectra always agrees with those of the spin-boson
model. This  establishes a mapping of our model at low 
temperatures onto an effective spin-boson model. This mapping becomes
intuitively clear by inspecting Fig.~\ref{fig:energy-parabel}(b). It 
shows a coupling between the two lowest states
$\ket{\pm J}$ on the parabola. For $T\to 0$, it is expected that the
model can be replaced by an effective spin-boson model with an
tunneling rate $\Delta$ which is a complicated function of
parameters $A$ and $B_{2n}$. Therefore, it is not surprising that our
calculations for the anisotropic large spin model reveal the
similarities to the results by Bulla et
al.~\cite{BullaVoita2005}, who established such a quantum phase
transition in the SBM.
Since the coupling of the two states
$\ket{\pm J}$ to the bosonic bath is proportional to $J_z$, we find
that  $\alpha_c(s) (2J)^2$  resembles the  phase boundary of
spin-boson model. The critical coupling $\alpha_c$ is proportional to
$1/J^2$.

\begin{figure}
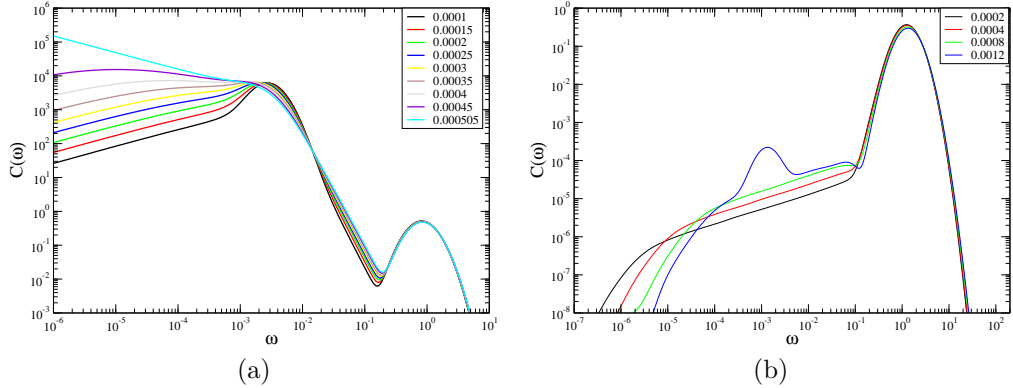

  \begin{tabular}{cc}
    \includegraphics[height=4.6cm,clip]{fig3a}
& \includegraphics[height=4.6cm,clip]{fig3b}
    \\
(a)&     (b) 
  \end{tabular}

  \centering

  \caption{Imaginary part of the spin-spin correlation
    function $C(\w)$ for different values of $\alpha$, (a) $J=3$
    and (b) $J=7/2$ in the
    sub-ohmic regime $s=0.5$ and    $T\to 0$. Parameters as in 
    Fig.~\ref{fig:phase-diagram}.
    }
\label{fig:chi-s=05}
\end{figure}

Such a behaviour is also observed in the equilibrium
spin-spin correlation function $C(\w)=\Im m \ll S_z | S_z\gg (\w-i\delta)$. 
The correlation function $C(\w)$ is calculated using the sum-rule conserving
algorithm \cite{PetersPruschkeAnders2006,WeichselbaumDelft2007}. We plot
$C(\w)$  on a log-log scale for a series of coupling constants and the
sub-ohmic regime $s=0.5$ in Fig.~\ref{fig:chi-s=05}(a).
The high energy feature stemming from the
easy-axis energy splitting proportional to $A$ is clearly visible. At
lowest coupling strength 
$\alpha=10^{-4}$, the second peak at $\w\approx 3\times 10^{-3}$ is well
pronounced, originating from the weak effective quantum tunneling
rate which is a function of parameters of $\H_{loc}$:  $A,B_{2n}$ as well
as $J$. For $\alpha<\alpha_c$,  $C(\w)$ vanishes according to
$C(\w) \propto \w^{s}$, while  $C(\w)$
diverges as $\w^{-s}$ for $\w\to 0$ and $\alpha>\alpha_c$.

\subsection{Half-integer Spin $J$}

\begin{figure}
  \centering
  \includegraphics[width=10cm,clip]{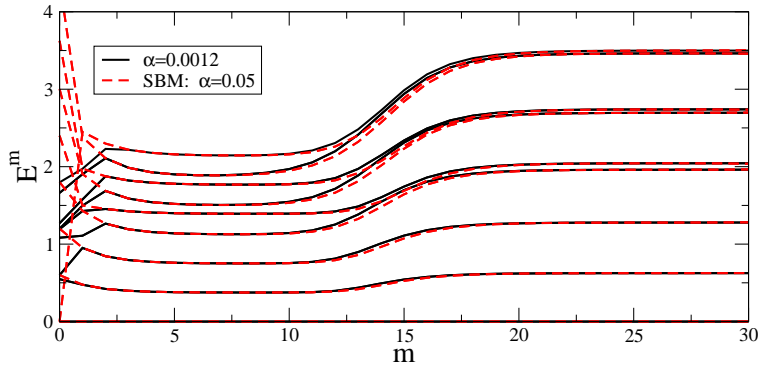}
  \caption{NRG level flow of the lowest eigenenergies $E_s^m$ 
    for $J=7/2$, $\alpha=0.0012$ in the 
    sub-ohmic regime $s=0.5$ in comparison with the level flow of the
    spin-boson model for $\Delta=0$ and $\alpha=0.05$. Parameters as in 
    Fig.~\ref{fig:phase-diagram}.
    }
\label{fig:level-flow-nj=8-sbm-D0}
\end{figure}

A completely different picture emerges for half-integer spins
$J$. Independent of the bath exponent $s$, no phase transition is 
found. As seen already  in Fig.~\ref{fig:energy-parabel}, the quantum
tunneling matrix elements $B_{2}$ and $B_{4}$ cannot  connect the two
states $\ket{\pm J}$.  The model can be mapped onto a  spin-boson
model with a vanishing tunneling rate $\Delta$ at low temperatures. To
illustrate this,  we compare the  NRG level
flow \cite{BullaCostiPruschke2007} of the large spin model
with the NRG level flow of a simple spin-boson model and $\Delta=0$.
The fixed-point spectra for $m\to\infty$ is identical for both
models as depicted in Fig.~\ref{fig:level-flow-nj=8-sbm-D0}. The
coupling constant  
$\alpha$ determines the crossover from high-temperature to the low
temperature fixed point: this crossover scale accidentally coincides
in both models for our choice of coupling constants. 
$C(\w)$ depicted in
Fig.~\ref{fig:chi-s=05}(b) is similar to the results of 
Fig.~\ref{fig:chi-s=05}(a) at high energies: the peak due to the
easy-axis splitting $A$ is also present. However, a
pronounced  peak at the effective quantum-tunnel energy is absent for
any half-integer $J$.

\section{Non-equilibrium dynamics}
\label{nonequi}
\begin{figure}
  \centering
  \includegraphics[width=10cm,clip]{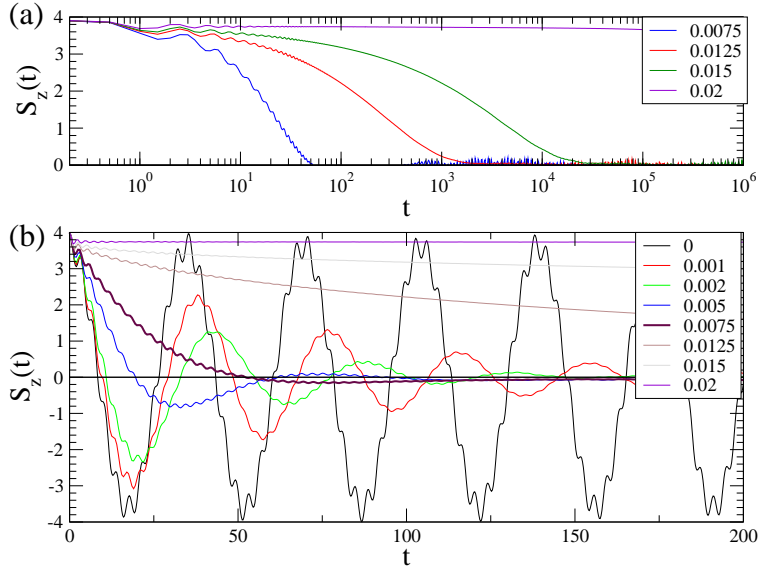}
  \caption{Real-time dynamics of the spin $S_z$ initially prepared in
    $S_z(0)=J=4$ for different values of $\alpha$ in the
    ohmic regime  $T = 3\times 10^{-5}$. Parameters: $A=0.1$, $B_2=10^{-3}$,
    $B_4=2 10^{-3}$, $\Lambda=2$, $N_s=160$, $N_b=8$, $N_z=16$.
    }
\label{fig:chi-s=1-nj=9}
\end{figure}

\begin{figure}
  \centering
  \includegraphics[width=12cm,clip]{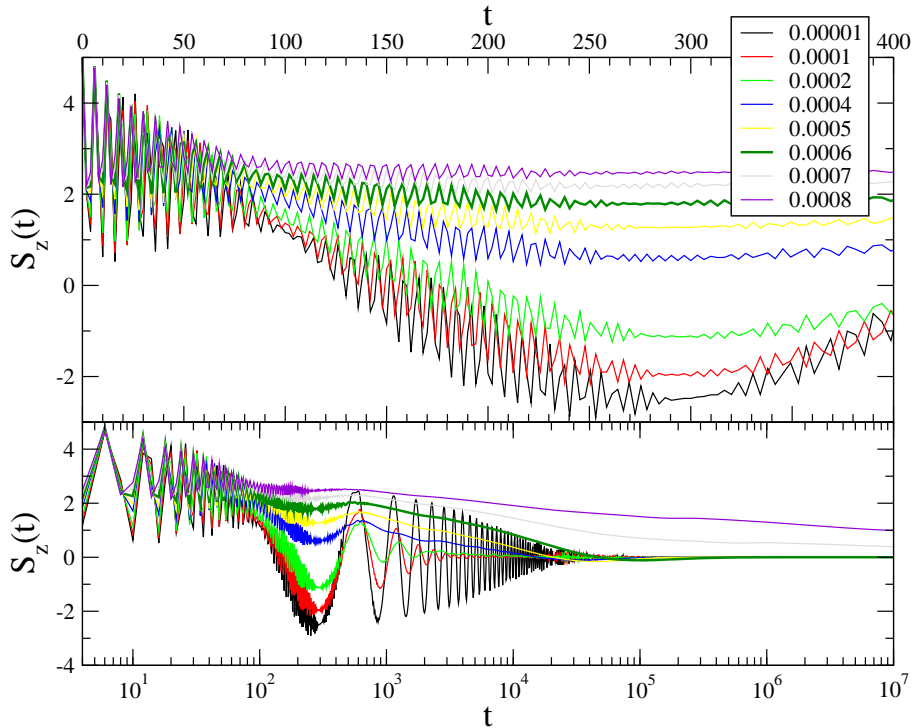}
  \caption{Real-time dynamics of the spin $S_z$ initially prepared in
    $S_z(0)=J=5$ for different values of $\alpha$ in the
    sub-ohmic regime  $T = 10^{-6}$. The upper panel (a) shows the short time
    dynamics on a linear time scale, while (b) resolves the real-time
    dynamics on a logarithmic time scale. For couplings
    $\alpha>\alpha_c\approx =0.00061$, the finite value of
    $S_z(t\to\infty)$ is a consequence of the quantum-phase transition
    into a localized phase. Parameters: $A=0.1$, $B_2=10^{-3}$,
    $B_4=2\times 10^{-3}$, $\Lambda=2$, $N_s=160$, $N_b=8$, $N_z=16$.
    }
\label{fig:chi-nj=11-s=05}
\end{figure}

For the investigation of the  non-equilibrium dynamics, the local
spin is prepared in a maximal polarized initial state with the
expectation value $S_z(0)=J$ by applying a strong local magnetic field
in $z$-direction. The external magnetic field is switched 
off at $t=0$.

Fig.~\ref{fig:chi-s=1-nj=9} displays the real-time dynamics of
$S_z(t)$ for a series of coupling strength $\alpha$ and an ohmic bath
($s=1)$, $J=4$ and the easy-axis splitting $A=0.1$. A weak
quantum-tunneling was enabled by $B_2=10^{-3}$ and $B_4=2\times
10^{-3}$. As a reference, the dynamics of a decouple spin ($\alpha=0$)
is included in the graph. For small $\alpha$,  damped coherent
oscillations with a superimposed short-time dynamics of small
amplitude are clearly visible. The oscillation frequency decreases
with increasing $\alpha$ and vanished at a finite coupling $\alpha$.
Above that value -- here $\alpha\approx 0.075$ -- only a spin decay is
observed. This corresponds to the regime $0.5<\alpha<1$ of the SBM. When
approaching  the critical coupling $\alpha_c(J=4)\approx 0.016$,
$S_z(t)$ only decays insignificantly on a 
very short time scale of the order $O(1)$ and remains constant for
$t\to\infty$, as seen in panel Fig.~\ref{fig:chi-s=1-nj=9}(a). This
behaviour occurs in the localized phase of the model where the effective tunneling rate is renormalized to zero.

A different picture emerges in the sub-ohmic regime. The
real-time dynamics of a large anisotropic spin  with $J=5$ is depicted
for different coupling strength in Fig.~\ref{fig:chi-nj=11-s=05}. The
upper panel shows the dynamics on a linear time axis. In the weak 
coupling regime, coherent high and low frequency oscillations are
clearly visible, in addition to a slow decay  of the amplitude. At a
critical coupling of $\alpha=0.00061$, we found a quantum-phase
transition in the equilibrium NRG calculations. As discussed in
Sec.~\ref{sec:qpt}, the quantum-critical point separates a delocalized
from a localized phase. As seen 
in Fig.~\ref{fig:chi-nj=11-s=05}, coherent oscillations prevail
in the short-time dynamics even in the localized phase close to the
quantum-phase transition. A finite time is required  before the
strong correlations can build up. In the long-time limit, $S_z(t)$ approaches
a finite value in the localized phase which should be proportional to
the order parameter for the QPT. These results are very similar to the
dynamics reported in the sub-ohmic spin-boson
model \cite{AndersBullaVojta2007}. In our case, larger number of
eigenstates of $H_{loc}$ yields  additional coherent oscillations
absent in the simple model. 

We expect that coherent oscillations prevail  to longer time-scales
with decreasing exponent $s$. Simultaneously, the critical coupling
$\alpha_c$ reduces strongly with deceasing exponent $s$, as  shown in
Fig.~\ref{fig:phase-diagram}.

\section{Decoherence}

\begin{figure}
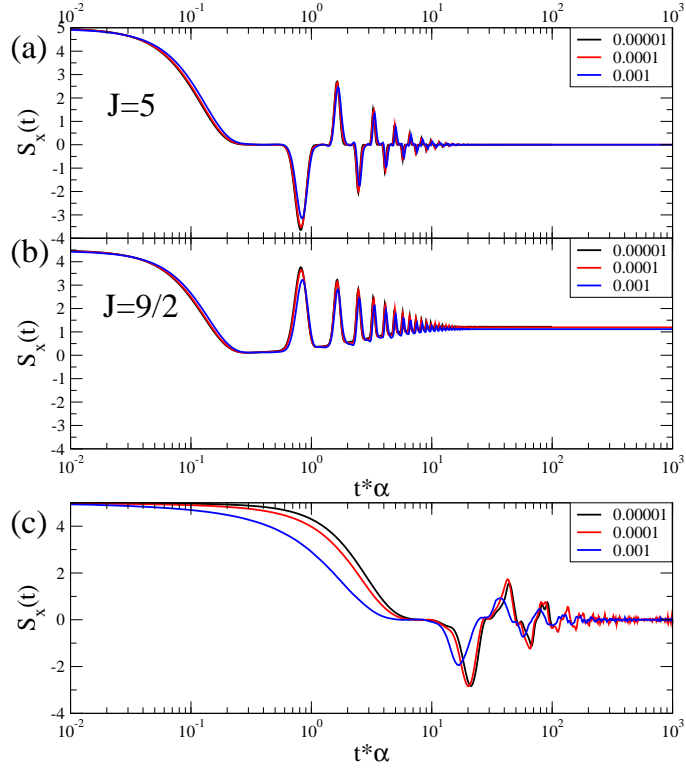

  \centering

  \includegraphics[width=90mm,clip]{fig7a}
  \includegraphics[width=90mm,clip]{fig7b}

  \caption{$S_x(t)$ vs $t*\alpha$ for three different coupling strength
    $\alpha=10^{-4},10^{-3},10^{-2}$ for $J=5$ (a) and
    $J=9/2$ (b) values of the spin in the $s=0.5$ sub-ohmic
    regime. $S_x(t)$ measures the     decoherence of the off-diagonal
    matrix elements of     $\rho_{i,j}^{red}$. Parameters:
    $T=10^{-6}$, $A=B_{2n}=0$.
    (c) shows $S_x(t)$ vs $t*\alpha$ for the same values of $\alpha$
    as (b) but with $A=2\alpha\omega_c/s$, $B_{2n}=0$. NRG parameters
    as in Fig.~\ref{fig:chi-s=1-nj=9}.
  } 
  \label{fig:decoherence}
\end{figure}

Decoherence is caused by the interaction of a subsystem with its
environment. The concept of
decoherence \cite{Zurek2003,Schlosshauer2005} provides inside into how a
coherence quantum state $\ket{s}$ of the subsystem is destroyed by the
entanglement with the environment. The reduced density matrix \cite{Feynman72}
\begin{equation}
  \rho_{i,j}^{red}(t) = \sum_{e} \bra{i,e}\hat \rho(t)\ket{j,e}
\end{equation}
evolves from a matrix describing such a  pure state $\ket{s}$ into a
matrix describing an ensemble. In an appropriate local basis $\ket{i}$,
the off-diagonal matrix elements of $\rho_{i,j}^{red}$ must
vanish for $t\to\infty$ \cite{Zurek2003,Schlosshauer2005}. In a
diagonally coupled  spin-boson model as described by the interaction $\H_I$,
Eq.~(\ref{eqn:H-I}), this would be the eigenstates of $S_z$
\cite{Zurek2003}. In the absence of quantum tunneling, $S_z$ also
commutes with $\H_{loc}$ and no energy is exchanged by the coupling
$\H_I$ to the environment. Decoherence is induced by the dephasing
with continuum of 
the bath modes each contributing a different phase shift. For
$J=1/2$, this is a well studied problem, and it was shown
analytically \cite{Unruh1995,PalmaSuominenEkert1996} that the
off-diagonal component of the reduced density matrix for the spin
can be written as $\rho_{\uparrow,\downarrow}(t) = e^{-\Gamma(t)}
\rho_{\uparrow,\downarrow}(0)$, where $\Gamma(t)$ is given by the exact
analytic expression
\begin{equation}
\Gamma(t) = \frac{1}{\pi}
            \int_{0}^\infty d\omega \, J(\omega)
                     \coth \left(
                                  \frac{\omega}{2T}
                           \right)
                     \frac{1-\cos(\omega t)}{\omega^2} \; .
\label{Gamma-via-J}
\end{equation}
Here $T$ is the temperature. Note that decoherence of
the pure quantum state $|s \rangle$ does not follow a
simple exponential form. Approximating the dephasing by a function
$\propto \exp(-t/T_{deph})$, which defines a single ``dephasing''
time-scale $T_{deph}$, might be insufficient even in very simple models
such as the SBM, since $\Gamma(t)$ cannot be replaced by $t/T_{deph}$.

We initially  prepared the spin system  in a pure state $\ket{s}$
comprising of a linear combination of all $S_z$ eigenstates 
\begin{equation}
  \ket{s} = \frac{1}{2J+1}\sum_{m=-J}^J \ket{m}
\end{equation}
by an appropriate choice of the initial Hamiltonian $\H_{loc}^i$.

We can gain some information on the decoherence
of a large anisotropic spin by measuring the time-dependent
expectation value $S_x(t)=\expect{\hat   S_x}(t)$, which only depend
on the off-diagonal matrix elements of $ \rho_{i,j}^{red}(t)$. The
quantum-tunneling parameters were set to $B_{2n}=0$ in the calculations.
Fig.~\ref{fig:decoherence} shows the dynamics of $S_x(t)$ on a
rescaled dimensionless  time axis $t\alpha$ for a
integer ($J=5$)   and a  half-integer ($J=9/2$) value of the spin in the
$s=0.5$ sub-ohmic regime.
In both cases, we observe an oscillatory decay of $S_x$ different from
the result for the spin-boson model, Eq.~(\ref{Gamma-via-J}). Here,
the finite coupling $\alpha$ introduces an effective $A$ via the
reorganization energy $E_{1g}$
\begin{eqnarray}
  E_{1g} = \sum_{q}\frac{\lambda_q^2}{\w_q} = \int_0^\infty d\w
  \,\frac{J(\w)}{\pi \w} = \frac{2\alpha\w_c}{s} \;\; , \;\; s>0
\punkt
\end{eqnarray}

The effective energy $\tilde E_m \propto(A -E_{1g}) m^2$ of
eigenstates $\ket{m}$ of $S_z$  induces quantum-oscillations whose
oscillation frequency is proportional to $\alpha$ for $A=0$. This $\alpha$
dependence of the frequency 
is revealed by plotting $S_x(t)$ versus the dimensionless timescale
$t\alpha$ so that the oscillation beats coincide and occur  on a time
scale of $t\alpha \approx O(1)$.  Since all eigenenergies  $\tilde E_m$ are
related to each other by integer values proportional to $A
-E_{1g}$, the local dynamics is characterized by a
recurrence time proportional to $1/(A -E_{1g})$ at which the original
local state would reappear in the absence of a spin-environment coupling.
However, the envelope function describing the time-dependence of the
recurrence amplitude decreases on a time scale proportional to
$1/\alpha$.

We can prolong this recurrence time by an appropriate counter-term in
the Hamiltonian. The curves of 
Fig.~\ref{fig:decoherence}(c) are obtained for the same parameters as
in Fig.~\ref{fig:decoherence}(a) and by setting $A=2\alpha\w_c/s$.  
Due to the discretization of the bath continuum, the cancellation is not
perfect, but the oscillation frequency is lowered by almost two
decades.

\section{Conclusion}

The equilibrium and non-equilibrium dynamics of an anisotropic large spin 
coupled to a dissipative sub-ohmic bosonic environment has been investigated
using the non-perturbative NRG. Such a large spin
might be realized in a single-molecular magnet. The coupling of the
bath modes to the easy-axis component of the spin enhances the
tendency for localization.  A quantum-phase transition exists only for
integer spin values  in the sub-ohmic regime, and the quantum-critical
point is associated with a new fixed point. These results generalized the work
of Bulla et al.~\cite{BullaBoson2003} for the spin-boson model. The
spin-spin correlation function $C(\w)$ diverges approximately as $|\w|^{-s}$
at and above the critical coupling $\alpha_c$.  The critical coupling
for $s\to 1$ is related to the critical coupling of the spin-boson
model by $\alpha_c=\alpha_c^{SBM}/(2J)^2$. We have shown
that the fixed point of the half-integer model is  identical to the
$S=1/2$  spin-boson model without quantum-tunneling $\Delta$. No
quantum-phase transition is found for half-integer $J$ due to the lack
of quantum-tunneling between the two states $\ket{\pm J}$.

We investigated the spin decay by switching off an external easy-axis
magnetic field at $t=0$. In the ohmic regime ($s=1$), the real-time
dynamics of $S_z(t)$ is governed by several local frequencies stemming
from the easy-axis splitting $A$ and the quantum-tunneling matrix
elements $B_{2n}$. With increasing coupling $\alpha$, the
quantum-tunneling is reduced, but the splitting $A$ is increased by
the reorganization energy $E_{1g}$. The low-frequency 
coherent oscillations vanish at some value $\alpha$ but the
high-frequency coherent oscillations prevail with a very small amplitude.
Further increasing of $\alpha$ beyond the critical coupling
$\alpha_c$ drives the system in the localized phase. Here, the spin
decays weakly on a very short time-scale of the order of the
reciprocal cutoff $1/\omega_c$.
A completely different picture emerges in a
sub-ohmic environment. Oscillatory response is found at short and
intermediate time-scales  even in the localized phase at coupling strength
$\alpha> \alpha_c$. 

In the calculation of the decoherence through dephasing, we identified
the occurring long-time oscillation frequency with recurrence
frequency which is dependent on the reorganization energy. These
oscillations can be reduced and suppressed by adding a counter-term in
the original Hamiltonian. 
We have demonstrated that the non-perturbative renormalization group
approach can describe such  equilibrium and non-equilibrium dynamics
close to a quantum-phase transition.

\ack
We acknowledge stimmulating discussions  with J.~Bartholomew,
R.~Bulla and S.~Tornow.
This research was supported in part by the DFG by project AN
275/6-1. We acknowledge supercomputer support by the NIC, 
Forschungszentrum J\"ulich under project no.\ HHB000.

\vspace*{5mm}


\providecommand{\newblock}{}

\end{document}